\documentclass[preprintnumbers,showkeys,floats,prd,aps,nofootinbib]{revtex4}
\usepackage{tipa}
\usepackage{amssymb,amsmath}
\usepackage[dvips]{graphicx}

\begin{document}

\title{Reconstructing a String-Inspired Quintom Model of Dark Energy}
\author{Shuai Zhang}\email{zhangshuai_237@126.com}
\author{Bin Chen}\email{bchen01@pku.edu.cn}
\affiliation{
  Department of Physics, and State Key Laboratory of Nuclear Physics and Technology,\\
  Peking University
  Beijing 100081, People¡¯s Republic of China
}
\date{June 27, 2008}

\begin{abstract}

In this paper, we develop a simple method for the reconstruction of
the string-inspired dark energy model with the lagrangian $ {\cal L}
= -V ( \phi ) \sqrt{ 1 - \alpha' {\nabla}_{\mu} \phi \nabla^{\mu}
\phi + \beta' \phi \Box \phi} $ given by Cai et. al., which may
allow the equation-of-state parameter cross the cosmological
constant boundary $ ( w=-1 ) $. We reconstruct this model in the
light of three forms of parametrization for dynamical dark energy.

\end{abstract}

 \preprint{}
\maketitle

\section {Introduction} \label{1}

A number of recent cosmological data, including supernovae(SN)
Ia~\cite{SN}, large-scale structure(LSS)~\cite{LSS}, and cosmic
microwave background (CMB) anisotropy~\cite{CMB}, have provided
strong evidences for a spatially flat and accelerated expanding
universe at the present time. In the context of
Friedmann-Robertson-Walker (FRW) cosmology, this acceleration is
attributed to the domination of a mysterious component, dubbed dark
energy (DE). Although a model of cosmological constant is consistent
with the observations, it suffers serious difficulties, for example,
the present value of the cosmological constant $\Lambda$ is
$10^{123}$ times smaller than that predicted by particle physics on
condition that the UV cutoff is at Plankscale, which is the
so-called coincidence problem.~\cite{Dynamics of dark energy} Hence,
as a possible solution to these problems various dynamical models of
DE have been proposed.

Up to now, a large number of scalar-field dark energy models has
been proposed, such as quintessence~\cite{quintessence},
K-essence~\cite{K-essence}, tachyon~\cite{tachyon},
phantom~\cite{phantom}, quintom~\cite{quintom} and ghost
condensate~\cite{ghost condensate1}. The analysis of the
observational data of type Ia supernova, mainly the 157 ``gold''
data listed in~\cite{SN}, shows that the equation-of-state (EoS) of
dark energy $w$ is likely to cross the cosmological constant
boundary $-1$. Nevertheless, the dark energy models with a single
scalar field with a lagrangian of form ${\cal L} = {\cal
L}(\phi,\partial_{\mu} \phi \partial^{\mu} \phi) $ such as
quintessence and phantom only allow $w$ to be either larger or
smaller than $-1$, according to the ``no-go'' theorem~\cite{no-go}.
This is why a class of dynamical models dubbed quintom has to
require two fields in order to permit EoS across $-1$.

In this paper, we will discuss a string-inspired quintom model of
dark energy proposed in Ref.~\cite{a string-inspired quintom model
of dark energy}, of which the action is given by
\begin{equation}\label{eq:action}
S = \int{d^4x\sqrt{-g} \left[ -V ( \phi ) \sqrt{ 1 - \alpha'
{\nabla}_{\mu} \phi \nabla^{\mu} \phi + \beta' \phi \Box \phi}
\right]}~.
\end{equation}
The signature $(+,-,-,-)$ is used in this paper. This model
generalizes the usual ``Born-Infeld-type'' action for the effective
description of tachyon dynamics by adding a term $ \phi\Box\phi $ to
the usual $ \nabla_{\mu}\phi\nabla^{\mu}\phi $ in the square root.
$V (\phi)$ is a potential of scalar field $\phi$ with dimension of
[mass]$^4$.

In a flat FRW spacetime, from the variation of the action
(\ref{eq:action}) with respect to $\phi$, we obtain the equations
of motion of the homogenous scalar $\phi$,
\begin{eqnarray}
\label{eq:motion1} \ddot{\phi}+3H\dot{\phi} &=& \frac{\beta V^2
\phi}{4M^4 \psi^2} -
\frac{M^4}{\beta\phi} + \frac{\alpha}{\beta}\dot{\phi}^2~,\\
\label{eq:motion2} \ddot{\psi}+3H\dot{\psi} &=&
(2\alpha+\beta)\left(\frac{M^4\psi}{\beta^2\phi^2} -
\frac{V^2}{4M^4\psi}\right) -
(2\alpha-\beta)\frac{\alpha\psi}{\beta^2\phi^2}\dot{\phi}^2-\frac{2
\alpha}{\beta \phi}\dot{\phi}\dot{\psi} - \frac{\beta V
\phi}{2M^4\psi}V_{\phi}~,
\end{eqnarray}
where we have defined the new variables $\psi$ and $f$
to simplify the calculations,
\begin{eqnarray}\label{eq:psi}
\psi &\equiv& \frac{\partial \cal L}{\partial \Box \phi} = -
\frac{V
\beta \phi}{2 M^4 f}~, \\
\label{eq:f} f &=& \sqrt{1 - \alpha' {\nabla}_{\mu} \phi
\nabla^{\mu} \phi + \beta' \phi \Box \phi}~.
\end{eqnarray}
Here it is defined that $\beta=M^4\beta'$, $\alpha=M^4\alpha'$,
with $\alpha$ and $\beta$ being dimensionless parameters, and $M$
an energy scale used to make the ``kinetic energy terms''
dimensionless. And the variation of the action (\ref{eq:action})
with respect to $g_{\mu \nu}$ gives the density and the pressure
of dark energy:
\begin{eqnarray}\label{eq:density}
\rho &=& Vf + \frac{d}{a^3 dt}(a^3 \psi \dot{\phi}) + \frac{\alpha
V}{M^4 f}\dot{\phi}^2 - 2\dot{\psi}\dot{\phi}~,\\
\label{eq:pressure} p &=& -Vf - \frac{d}{a^3 dt}(a^3 \psi
\dot{\phi})~,
\end{eqnarray}
then we obtain
\begin{equation}\label{eq:plus}
\rho+p = \frac{\alpha V}{M^4 f}\dot{\phi}^2 -
2\dot{\psi}\dot{\phi}~.
\end{equation}
As is discussed in Ref.~\cite{a string-inspired quintom model of
dark energy}, $w$ approaches $-1$ and then crosses over it  on
condition that (i)$ \dot{\phi}=0 ,~ \phi \neq 0 ,~ \ddot{\phi}\neq 0
,~ \frac{d}{dt} \Box \phi \neq 0$ or (ii) $ \phi \neq 0 ,~
\dot{\phi} \neq 0 ,~ Y=0 ,~ \dot{Y}\neq 0 $, where $ Y = (\alpha +
\beta)\frac{V}{f}\dot{\phi}+\beta \phi
\frac{d}{dt}\left(\frac{V}{f}\right)$.

In this paper, we will reconstruct this model in the light of three forms of parametrization. Next, in section \ref{2}, we present our program to reconstruct the model. In section \ref{3} from three forms of
parametrization of EoS, which fit observational data of the joint
analysis of SNIa+CMB+LSS, we reconstruct the model. Finally, in Section \ref{4}, we end with a  short summary of our results and some discussions.

\section {Reconstruction Program} \label{2}

In this section, we shall perform a reconstruction program of this
model. The aim of this section is to express the physical
quantities to be reconstructed in respect of redshift $z$. So once
the form of parametrization of EoS is given, we can directly work
out the relationships between any two of the physical quantities,
including that of the potential $V$ and the scalar field $\phi$,
which is what we expect eventually.

In terms of $\psi$, we can rewrite
\begin{eqnarray}\label{eq:rho}
\rho &=& -\frac{\beta V^2 \phi}{4 M^4 \psi}-\frac{\alpha \psi
\dot{\phi}^2}{\beta \phi}-\dot{\psi}\dot{\phi}-\frac{M^4
\psi}{\beta \phi}=\tilde{K}+\tilde{V}~,\\
\label{eq:pl} \rho+p &=& -2 \frac{\alpha \psi \dot{\phi}^2}{\beta
\phi}-2 \dot{\psi}\dot{\phi}=2 \tilde{K}~,
\end{eqnarray}
where the effective kinetic energy term $\tilde{K}$ and the
effective potential term $\tilde{V}$ are defined as
\begin{eqnarray}\label{eq:K}
\tilde{K} &=& -\frac{\alpha \psi \dot{\phi}^2}{\beta \phi}-
\dot{\psi}\dot{\phi}~,\\
\label{eq:V} \tilde{V} &=& -\frac{\beta V^2 \phi}{4 M^4
\psi}-\frac{M^4 \psi}{\beta \phi}~.
\end{eqnarray}
Therefore, the Friedmann equations in a flat FRW spacetime are
given by
\begin{eqnarray}
\label{eq:H1} 3 M_{pl}^2 H^2&=&\rho_{m}+\rho
~=~ \rho_{m}+\tilde{K}+\tilde{V}~, \\
\label{eq:H2} 2 M_{pl}^2 \dot{H} &=& -\rho_{m}-\rho-p ~=~
-\rho_{m}-2\tilde{K},
\end{eqnarray}
where $M_{pl}^2 \equiv (8 \pi G)^{-1/2}$ is the reduced Planck
mass, and $\rho_{m}$ is the energy density of dust matter. The
EoS of dark energy is
\begin{equation}\label{eq:EoS}
w\equiv\frac{p}{\rho}=-1+\frac{2}{1+\tilde{V} / \tilde{K}}.
\end{equation}
From Eq. (\ref{eq:EoS}), it is obviously that $w>-1$, when $
\frac{\tilde{V}}{\tilde{K}}>-1$, while $w<-1$, when
$\frac{\tilde{V}}{\tilde{K}}<-1$. And when
$\tilde{V}+\tilde{K}=0$, the transition takes place.

From Eq. (\ref{eq:H1}) and Eq. (\ref{eq:H2}), we can easily get
\begin{equation}\label{eq:V''}
\tilde{V}=-\frac{1}{2} \rho_{m}+3 M_{pl}^2 H^2+M_{pl}^2 \dot{H},
\end{equation}
\begin{equation}\label{eq:K''}
\tilde{K}=-\frac{1}{2} \rho_{m}-M_{pl}^2 \dot{H}.
\end{equation}
As in our model, the dark energy fluid does not couple to the
background fluid, the expression of the energy density of dust
matter in respect of redshift $z$ is
\begin{equation}\label{eq:density-m}
\rho_{m}=3 M_{pl}^2 H_{0}^2 \Omega_{m0} (1+z)^3,
\end{equation}
where $\Omega_{m}$ represent the ratio density parameter of matter
fluid and the subscript ``0'' indicates the present value of the
corresponding quantity. By using the relationship
\begin{equation}\label{eq:dz-dt}
\frac{d}{dt}=-H(1+z)\frac{d}{dz},
\end{equation}
$\tilde{V}$ and $\tilde{K}$ in respect of $z$ can be expressed as
\begin{eqnarray}
\tilde{V}(z) &&=-\frac{3}{2} M_{pl}^2 H_{0}^2
\Omega_{m0}(1+z)^3-\frac{1}{2} M_{pl}^2 H_{0}^2 r'(1+z)+3M_{pl}^2
H_{0}^2 r,
\label{eq:V'}\\
\tilde{K}(z) &&=-\frac{3}{2} M_{pl}^2 H_{0}^2
\Omega_{m0}(1+z)^3+\frac{1}{2} M_{pl}^2 H_{0}^2 r'(1+z)
\label{eq:K'}
\end{eqnarray}
 where
\begin{equation}\label{eq:r}
r \equiv \frac{H^2}{H_{0}^2}.
\end{equation}

In this manner, once the expression of $r$ is given in respect of
$z$, $\tilde{V}(z)$ and $\tilde{K}(z)$ can be reconstructed from Eq.
(\ref{eq:V'}) and Eq. (\ref{eq:K'}). And by using Eq. (\ref{eq:K})
and Eq.~(\ref{eq:V}), $\tilde{K}$ and $\tilde{V}$ can also be expressed in respect
of $\phi$ or $\psi$, on condition that another relationship among
$\phi$, $\psi$ and $z$ is given. And it is not hard for us to get
the very relationship from equations (\ref{eq:motion1}),
(\ref{eq:V}), (\ref{eq:dz-dt}) and (\ref{eq:r}), which is
\begin{equation}
\phi'' r H_{0}^2 (1+z)^2 -2\phi' r H_{0}^2 (1+z) +
\frac{1}{2}r'H_{0}^2 (1+z)^2 \phi' + 2 \frac{M^4}{\beta
\phi}-\frac{\alpha}{\beta}\phi'^2 rH_{0}^2 (1+z)^2 +
\frac{\tilde{V}}{\psi}=0.
\label{eq:phi-psi}\\
\end{equation}

Now it is explicit that the expression of $\tilde{V}$ and
$\tilde{K}$ in respect of $z$ are the same as those
in~\cite{Hessence}. So the reconstructing procedure will be
similar to that in~\cite{Hessence}, except that the expression of
$\tilde{V}$ in respect of the fields $\phi$ and $\psi$ will be
given by the inverse function of Eq. (\ref{eq:V})
\begin{equation}\label{eq:inverse}
V^2 = -\frac{4M^4 \psi}{\beta\phi} \tilde{V}-\frac{4M^8
\psi^2}{\beta^2 \phi^2},
\end{equation}
and the expression of $\phi$ in respect of z is also different from that in \cite{Hessence}.
It will never escape our notice that as $V^2>0$, Eq.
(\ref{eq:inverse}) may restrict $\phi$ within an interval other
than the whole real number set. Next, in the following section, we
will obtain the relationship between the potential $V$ and the
scalar field $\phi$ in the light of three parametrizations of
$w(z)$(or $r(z)$, as $r(z)$ and $w(z)$ are related by Eq.
(\ref{eq:w}) and Eq. (\ref{eq:r-w})).

\section {Reconstruction Results} \label {3}

As has been discussed above, the two examples considered in the
third section of ~\cite{Hessence}, where parameterizations for
$r(z)$ is fitted to the latest 182 SNe Ia Gold dataset~\cite{182},
are also suitable for this model. In this paper, to provide the base
for the reconstruction of the string-inspired quintom model of dark
energy, we will use  three forms of parametrization that has been
discussed by Gong and Wang~\cite{parameter}. These forms of parametrization have also been
used in~\cite{ghost condensate}, and two of which is the very two
forms used in~\cite{Hessence}. The final results are derived from these
three parametrizations.

The three forms of parametrization are investigated uniformly here,
in order that it is convenient for us to compare one with another.
The three forms are 
\begin{itemize}
\item Parametrization 1:\begin{equation}\label{eq:parameter1}
w(z) = w_{0} + \frac{w_{a}z}{(1+z)},
\end{equation}
which was suggested by Chevallier $\&$ Polarski~\cite{C} and Linder~\cite{L},
to avoid the divergence problem effectively;
\item Parametrization 2:\begin{equation}\label{eq:parameter2}
w(z) = w_{0} + \frac{w_{b}z}{(1+z)^2},
\end{equation}
which was suggested by Jassal, Bagla $\&$ Padmanabhan~\cite{J};
\item Parametrization 3:\begin{equation}\label{eq:parameter3}
r(z) = \Omega_{m0}(1+z)^3+A_{0}+A_{1}(1+ z) + A_{2}(1 + z)^2,
\end{equation}
which was suggested by Alam, Sahni, Saini $\&$ Starobinsky~\cite{S}. It is
an interpolating fit for $r(z)$ having the right behavior for both
small and large redshifts.
\end{itemize}
The similar results for the ``parametrization 1'' and ``parametrization
3'' were also independently obtained in Ref.~\cite{N} and in
Ref.~\cite{A} respectively.

Using the new measurement of the CMB shift parameter~\cite{W},
together with LSS data (the BAO measurement from the SDSS
LRGs)~\cite{E} and SNIa data (182 ``gold'' data released
recently)~\cite{182}, the authors of Ref.~\cite{parameter}
constrained the parameters, whose fit results are summarized as
follows: for ``parametrization 1'', $\Omega_{\rm m0}=0.29\pm 0.04$,
$w_0=-1.07^{+0.33}_{-0.28}$ and $w_a=0.85^{+0.61}_{-1.38}$; for
``parametrization 2'', $\Omega_{\rm m0}=0.28^{+0.04}_{-0.03}$,
$w_0=-1.37^{+0.58}_{-0.57}$ and $w_b=3.39^{+3.51}_{-3.93}$; for
``parametrization 3'', $\Omega_{\rm m0}=0.30\pm 0.04$,
$A_1=-0.48^{+1.36}_{-1.47}$ and $A_2=0.25^{+0.52}_{-0.45}$.

From Eq. (\ref{eq:H1}), Eq. (\ref{eq:H2}) and Eq.
(\ref{eq:density-m}) we obtain the model-independent expression of
the EoS $w$ and the deceleration parameter $q$ :
\begin{eqnarray}
w(z) &&= \frac{p}{\rho} = \frac{(1+z)r'-3r}{3r-3\Omega_{m0}
(1+z)^3},
\label{eq:w} \\
r(z) &&= \Omega_{m0} (1+z)^3 + (1-\Omega_{m0}) \exp \left(3
\int_{0}^{z} \frac{1+w(\tilde{z})}{1+\tilde{z}}d\tilde{z}\right),
\label{eq:r-w}
\\
q(z) &&= -1-\frac{\dot{H}}{H^2}=-1-\frac{r'(1+z)}{2r}. \label{eq:q}
\end{eqnarray}
\begin{figure}[here]
\begin{center}
  \includegraphics{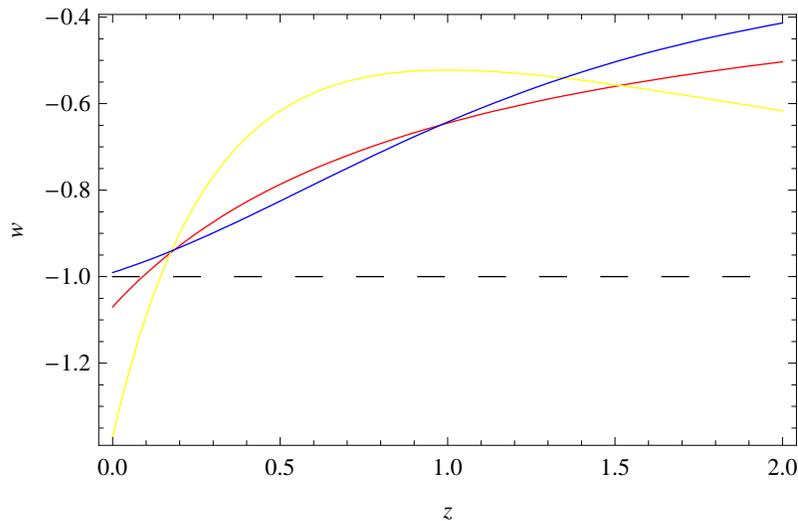}
  \caption{
    The evolutions of the EoS in respect of redshift
    $z$. The red, yellow and blue solid line represent parametrization 1, parametrization
    2, and parametrization 3 respectively. The best-fit values of the joint analysis of
    SNIa+CMB+LSS~\cite{parameter} is accepted here, namely, for
parametrization 1, $\Omega_{\rm m0}=0.29$, $w_0=-1.07$ and
$w_a=0.85$; for parametrization 2, $\Omega_{\rm m0}=0.28$,
$w_0=-1.37$ and $w_b=3.39$; for parametrization 3, $\Omega_{\rm
m0}=0.30$, $A_1=-0.48$ and $A_2=0.25$.
  }
\label{fig:w}
\end{center}
\end{figure}
\begin{figure}[h]
\begin{center}
  \includegraphics{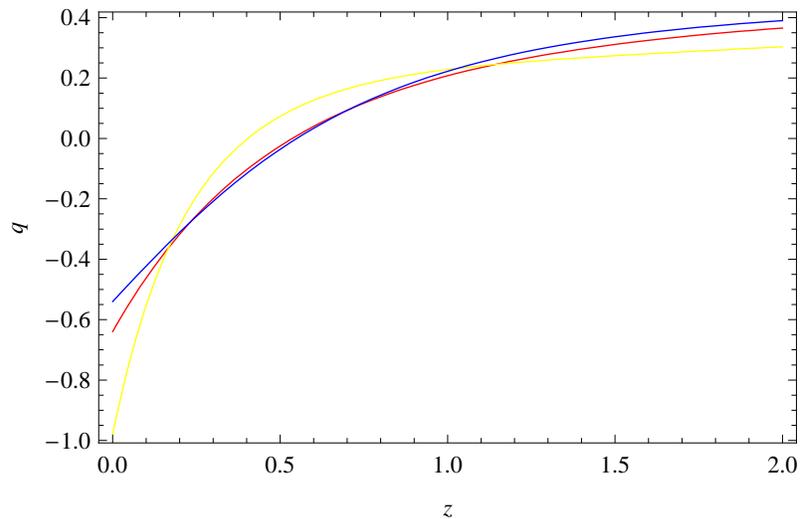}
  \caption{
    The evolutions of the deceleration parameter in respect of redshift
    $z$. The red, yellow and blue solid line represent parametrization 1, parametrization
    2, and parametrization 3 respectively. The best-fit values of the joint analysis of
    SNIa+CMB+LSS is accepted here.
  }
  \label{fig:q}
\end{center}
\end{figure}
The evolution of $w(z)$ and $q(z)$ are plotted in Fig.~\ref{fig:w} and  Fig.~\ref{fig:q} respectively. In Fig.~\ref{fig:w} we find that all
the three forms of parametrization require a model that permits
EoS to cross cosmological constant boundary $ ( w=-1 ) $. And
Fig.~\ref{fig:q} tells us our universe is experiencing an
accelerating expansion now.

\begin{figure}[h]
\begin{center}
  \includegraphics{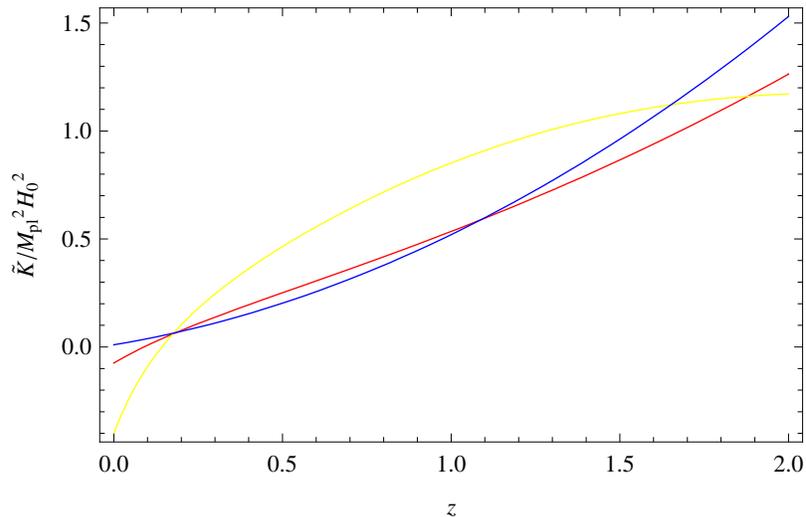}
  \caption{
    The reconstructed $\tilde{K}$ in respect of $z$. The red, yellow and blue solid line represent parametrization 1, parametrization
    2, and parametrization 3 respectively. The best-fit values of the joint analysis of
    SNIa+CMB+LSS is accepted here.
  }
  \label{fig:K}
\end{center}
\end{figure}

\begin{figure}[h]
\begin{center}
  \includegraphics{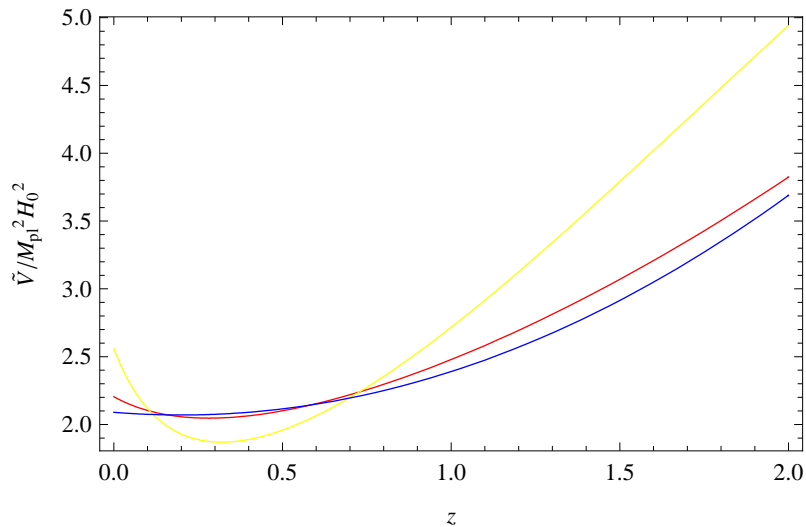}
  \caption{
    The reconstructed $\tilde{V}$ in respect of $z$. The red, yellow and blue solid line represent parametrization 1, parametrization
    2, and parametrization 3 respectively. The best-fit values of the joint analysis of
    SNIa+CMB+LSS is accepted here.
  }
  \label{fig:V}
\end{center}
\end{figure}

\begin{figure}[h]
\begin{center}
  \includegraphics{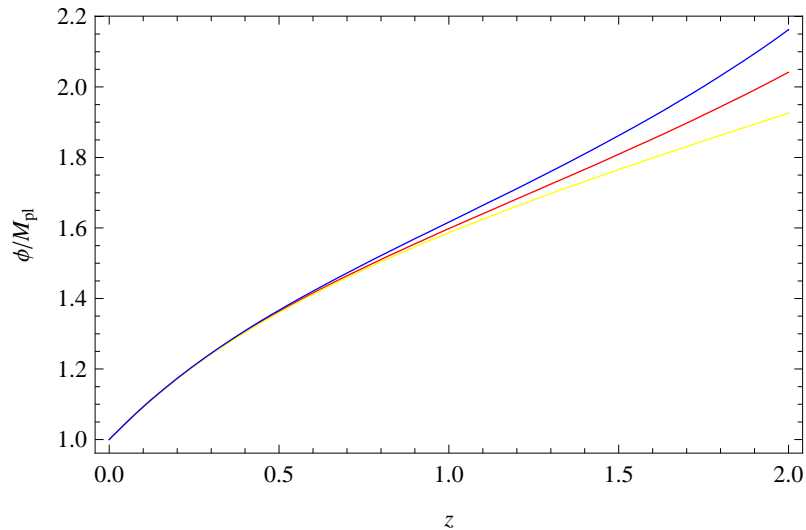}
  \caption{
    The reconstructed scalar field $\phi$ in respect of $z$.The red, yellow and blue solid line represent parametrization 1, parametrization
    2, and parametrization 3 respectively. The best-fit values of the joint analysis of
    SNIa+CMB+LSS is accepted here. And we set $\alpha=\beta=M=1$, $\psi(0)=\phi(0)=1$ and
    $\phi'(0)=1$.
  }
  \label{fig:phi}
\end{center}
\end{figure}

\begin{figure}[h]
\begin{center}
  \includegraphics{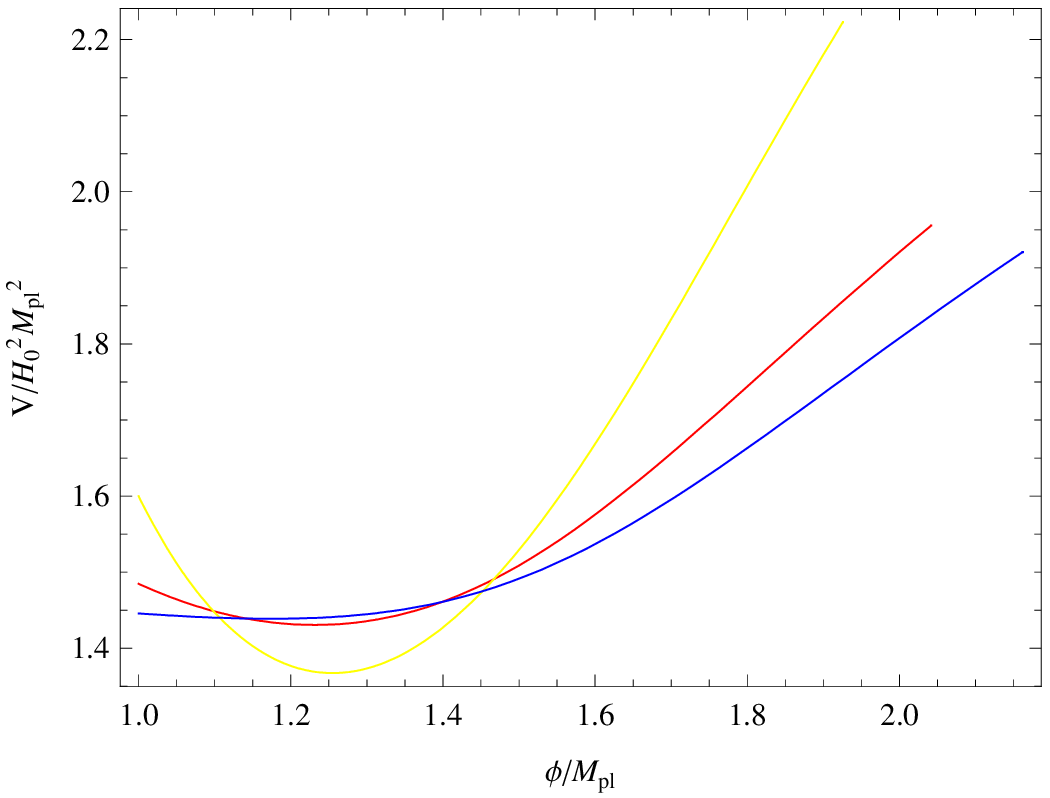}
    \caption{
    The reconstructed potential $V$ in respect of $\phi$.The red, yellow and blue solid line represent parametrization 1, parametrization
    2, and parametrization 3 respectively. The best-fit values of the joint analysis of
    SNIa+CMB+LSS is accepted here.
  }
  \label{fig:Vphi}
\end{center}
\end{figure}

According to Eq. (\ref{eq:V'}), Eq. (\ref{eq:K'}) and the three
parametrizations, the evolutions of $\tilde{K}(z)$ and
$\tilde{V}(z)$ are shown in Fig.~\ref{fig:K} and Fig.~\ref{fig:V}
respectively.
 As $\tilde{K}(z)$ and
$\tilde{V}(z)$ are not unknown any more, by solving the
differential equations (\ref{eq:K}) and (\ref{eq:phi-psi}), we get
the evolution of $\phi$ in respect of $z$, which is plotted in
Fig.~\ref{fig:phi}.
 Since now that $V$ and $\phi$ are all related to
$z$, we can directly obtain the relationship between the potential
$V$ and the scalar field $\phi$, which is plotted in
Fig.~\ref{fig:Vphi}.
Here we notice that some range of $\phi$ is forbidden by Eq.
(\ref{eq:inverse}).

By now, we have completed reconstructing the form of the potential
$V(\phi)$ of our model on the basis of observational data. The
result also tells us that this model is a viable candidate to
describe the dynamics of dark energy.

\section {Conclusions and Discussions} \label{4}

As is discussed in the first section, we need a model consisted of
two scalar fields to describe the dynamics of dark energy to make
it possible that EoS may cross the cosmological-boundary $w=-1$.
However, a general quintom model with double fields
\cite{Cai:2006dm} is not easy to be reconstructed, because there
is three physical quantities to be expressed in respect of
redshift $z$, which is the potential $V$, the scalar fields $\phi$
and $\psi$, while only two equations are viable, namely Friedmann
equations. But for a quintom with the potential related to only
one of the two fields and has nothing to do with another, such as
Hessence proposed in Ref.~\cite{Hessence'} and the string-inspired
model that we have just discussed, one of the equations of motion
does not contain derivative terms of $V$ with respect of $\phi$ or
$\psi$. So the very equation gives another relationship among $V$,
$\phi$ and $\psi$. As a result, it is possible for us to
reconstruct such models according to the observational data.

In fact, to some extent, we must confess that the quintom with a
potential related to only one scalar field is more or less a
``pseudo quintom'', because we can always represent one field with
an expression of another field by solving some equations. Our
model and Hessence (See Eq.~(12) in Ref.~\cite{Hessence}) are both
examples. And in this way, the lagrangian can be expressed with
just one scalar field. As a matter of fact, at the beginning of
this paper, Eq.~(\ref{eq:action}) does express the lagrangian with
only one field. It is Eq.~(\ref{eq:psi}) that defines a new scalar
field $\psi$ and make our model a quintom model, but, at the same
time, it gives the relationship between $\psi$ and $\phi$, which
makes the model possible to be reconstructed. We notice that
``no-go'' theorem just forbids the dark energy models with a
lagrangian of form ${\cal L} = {\cal L}(\phi,\partial_{\mu}
\partial^{\mu} \phi) $ from crossing $w=-1$. But Eq.~(\ref{eq:action}) does
not belong to this class, for it has the term $\beta' \phi \Box \phi$ in
the square root.

Finally, it is possible for a reconstructable quintom model to be
expressed as a model constructed by a single field. A
string-inspired quintom has provided a good example. From the
analysis of this paper, we find that the result of the
reconstruction is able to give us the expressions of $\phi$ and
$\psi$ in respect of $z$, which result in $\psi=\psi(\phi)$ and so
will eliminates one of the fields in the action. In this sense, we
might say a quintom model that fits the observational data is
equivalent to a model consisted of one scalar field with a
lagrangian other than ${\cal L} = {\cal L}(\phi,\partial_{\mu}
\partial^{\mu} \phi) $ that allows $w$ to cross $-1$.

\section*{Acknowledgments}
This work was partially supported by NSFC. Grant.\ No.\ 10535060,
10775002.


\end{document}